**Virial equation-of-state for the hard-disk fluid**

Leslie V Woodcock

Department of Chemical and Biomolecular Engineering
National University of Singapore, Singapore 117576
e-mail chewlv@nus.edu.sg

**Virial coefficients for the two-dimensional hard-disk fluid, when expressed in powers of density relative to maximum close packing ($\rho_0$), lead to an accurate closed equation-of-state for the equilibrium fluid, analogous to hard spheres [1]. The 2D series also converges for all densities up to a negative pole at $\rho_0$. The virial pressure begins to deviate from the thermodynamic fluid on approach of the ordering transition.**

In a recent publication [1], a closed equation-of-state for hard spheres from the virial expansion in powers of density relative to close-packing has been proposed. For the pressure of the hard-sphere fluid this virial expansion is

$$Z = 1 + \sum_n B_n (\rho/\rho_0)^{(n-1)} \qquad (1)$$

where $Z = pV/Nk_BT$: $k_B$ is Boltzmann's constant, $B_n$ are the dimensionless coefficients $\rho$ is the number density (N/V), and $\rho_0$ is the crystal close packing density. Equation (1) was shown to lead to a closed-form equation which is extremely accurate for the hard-sphere fluid (~100 times more accurate, for example, at fluid freezing, than the widely used Carnahan-Starling equation).

Here we derive a similar equation-of-state for hard disks. For 2D we use the same nomenclature except now $\rho = N/A$, A being the area, and $\rho_0 \sigma^2 = 2/3^{½}$ corresponding for the hexagonal close-packed 2D lattice structure. The literature recommended values for all the known virial coefficients for hard disks according to Kolafa and Rottner [2] are reproduced here in TABLE I. They express the virial coefficients in an alternative expansion in powers of the packing fraction (y)

$$Z = 1 + \sum_n \beta_n y^{(n-1)} \qquad (2)$$

where $Z = pA/Nk_BT$ and $\beta_n$ is related to $B_n$ in 2D by $B_n = \beta_n (\pi\rho_0/4)^{n-1}$

Also given in TABLE I are the values of the virial coefficients expressed in powers of the density relative to maximum close packing $\rho_0$ in equation (1).

From an inspection of all three dimensions, before the recently determined higher coefficients (i.e. > $B_8$) were known, it appeared that a simple closed-form equation-of-state might be possible on the observation that the virial coefficients in equation (1) were approaching $D^2$, i.e. 9 in 3D, and 4 in 2D (all $B_n = 1$ in 1D), i.e. the dimension squared. In both 3D and 2D however, the recent values $B_8$ to $B_{10}$ show that this is not the case. Nevertheless, it's interesting that in 2D and 3D, this expansion peaks close to $D^2$, which is the same as Nc –D, Nc being the crystal coordination number at $\rho_0$. We also note a similarity between D=2 and D=3 in that each has a single fluid-crystal phase transition. D=1 has no phase transition. D>3 show multiple crystal-crystal phase transitions. These observations are indicative of reflection of $\rho_0$ in the virial expansion .

TABLE I   Virial coefficients of the 2D hard-disk fluid: the values ($\beta_n$ in y-expansion) are taken from Kolafa and Rottner [2], the middle column gives the fitted values of $\beta_{12}$ to $\beta_{16}$ as predicted by the best-fit linear equation (3); the third column gives the same coefficients but as $B_n$ in the $\rho_0$-expansion, equation (1).

| n  | $\beta_n$ [2]   | $\beta_n$ [eqn.(3)] | $B_n$    |
|----|-----------------|---------------------|----------|
| 2  | 2               |                     | 1.813799 |
| 3  | 3.128017752     |                     | 2.572691 |
| 4  | 4.257854456     |                     | 3.175912 |
| 5  | 5.33689664      |                     | 3.610153 |
| 6  | 6.3630259       |                     | 3.903550 |
| 7  | 7.352077        |                     | 4.090396 |
| 8  | 8.318677        |                     | 4.197288 |
| 9  | 9.27234         |                     | 4.242903 |
| 10 | 10.2161         |                     | 4.23953  |
| 11 | 11.172          |                     | 4.2045   |
| 12 | 12.132± 0.03    | 12.15               | 4.1407   |
| 13 | 13.097± 0.06    | 13.04               | 4.0539   |
| 14 | 14.053± 0.08    | 13.98               | 3.945    |
| 15 | 14.94± 0.21     | 14.97               | 3.803    |
| 16 | 15.7± 0.4       | 16.02               | 3.62     |

Additional to the computed virial coefficients, using accurate MD data for disks, over the whole density range up to freezing, Kolafa and Rottner also determine the higher coefficients $B_{12}$ to $B_{16}$. These values are also given in TABLE I together with their calculated margins of uncertainty.

Inspection of virial coefficients plotted in powers of density relative to crystal close packing as in equation 1; (Fig. 2) shows that beyond $B_{11}$ they decrease approximately linearly according to

$$B_n = C - An \quad (n > 11) \qquad (3)$$

The $B_{12}$ to $B_{16}$ values of Kolafa and Rottner can be fitted to equation (3) with an EXCEL trendline regression $R^2 = 0.99$. the values for the two constants for the hard disk system are C =5.4995 and A = 0.1125., Equation (3 ), using these parameters, reproduces all the values $B_{12}$ to $B_{16}$ as shown in TABLE I to within the uncertainties quoted by Kalafa and Rottner [2].

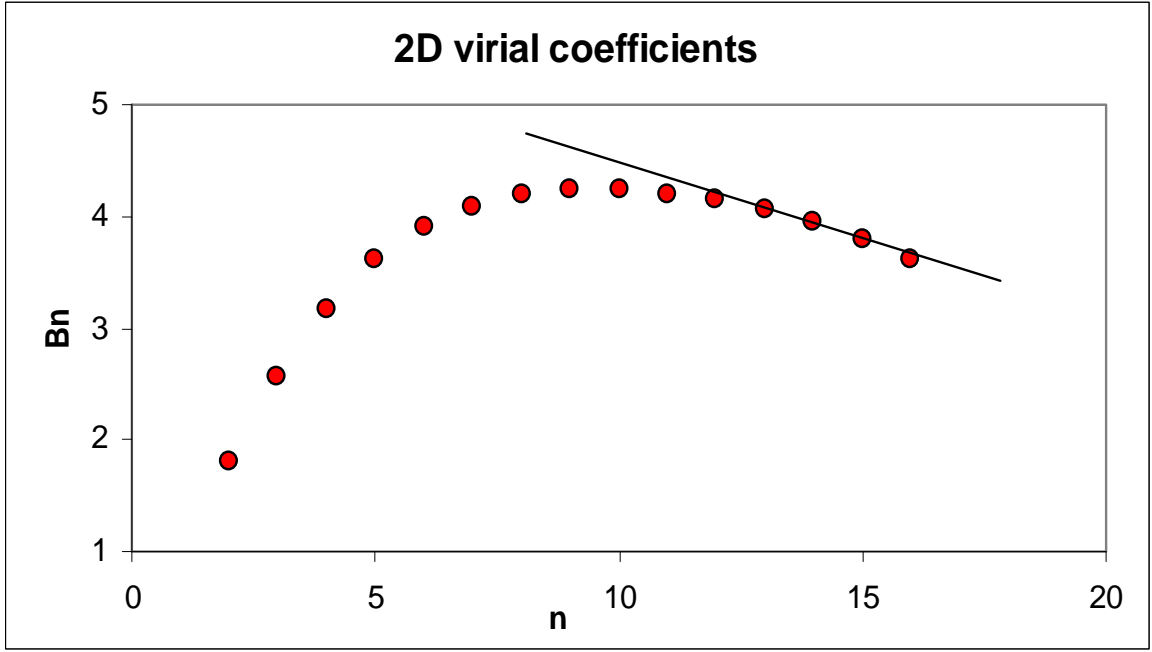

FIG.1. Virial coefficients of the hard-disk fluid in the expansion in powers of density relative to crystal close-packing: $B_2$- $B_4$ are analytic, $B_5$-$B_{10}$ are computed, $B_{11}$- $B_{16}$ were fitted to MD data by Kolafa and Rottner [1].

As is the case in 3D [1], we can hypothesize that if all the higher virial coefficients greater than $B_{11}$ can be obtained from equation (3). It is predicted that the virial coefficients become negative around $B_{50}$, and stay negative. This implies a virial equation-of-state that is continuous in all derivatives, and everywhere convergent, that itself eventually goes negative, and stays negative up to a pole at the maximum density $\rho_0$

The analytic closed-form equation-of-state for the hard-disk fluid, up to the density of the first phase transition, now takes exactly the same form in 2D as in 3D. If we let m denote the highest virial coefficient that is not given by equation (3), the same closed-form expression, for the summation of all powers of ρ greater than m, is obtained (see APPENDIX) .

$$Z = 1 + \sum_{n=2}^{m} B_n \rho^{*(n-1)} + \rho^{*m} \left[ \frac{(C-mA)}{(1-\rho^*)} - \frac{A}{(1-\rho^*)^2} \right] \quad (4)$$

where $\rho^* = \rho/\rho_0$ ; the EXCEL trendline values for the two constants C and A for the 2D hard-disk system are C = 5.4995 and A = 0.1125 and m = 11.

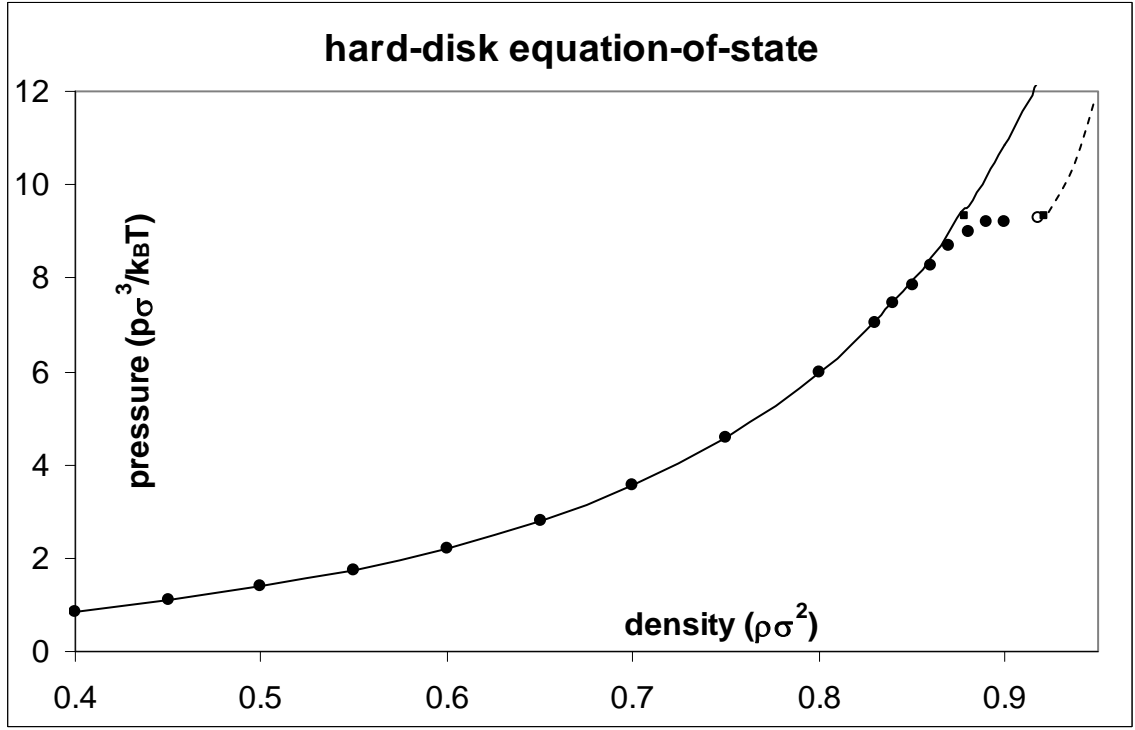

FIG.2 Equation-of-state for the hard-disk fluid in the high-density region from equation (4) in the text (solid line) compared with the MD data points of Kolafa and Rottner reference [2] (solid circles). Also plotted (open circle) is the state point from Jaster (reference [3]) for N= $(1024)^2$ alongside the Hoover-Ree [4] fluid freezing and crystal melting points (plotted as two solid squares at their calculated coexistence pressure 9.33)

The closed virial equation-of-state, equation (4) is seen from the comparisons with the MD data (TABLE II, FIG. 2 and FIG. 3) to be extremely accurate up to the vicinity of the phase transition, whereupon it begins to deviate. Whether the transition is either first order [4], or second order [3], it seems likely that the thermodynamic system may be exhibiting "pre-transition" effects which are commonly associated with weak first-order or second-order phase transitions, but not reflected in the virial expansion.

TABLE II.   Equation-of-state data for the hard-disk fluid given by equation (4) in the high-density fluid region, compared with the MD simulation data ( from reference [2])

| $\rho\sigma^2$ | 0.40 | 0.45 | 0.50 | 0.55 | 0.60 | 0.65 | 0.70 | 0.75 |
|---|---|---|---|---|---|---|---|---|
| MD | 2.151439 | 2.427668 | 2.760123 | 3.164787 | 3.663691 | 4.287926 | 5.082362 | 6.113391 |
| Eq. 4 | 2.151443 | 2.427667 | 2.760123 | 3.164788 | 3.663699 | 4.287942 | 5.082536 | 6.114721 |

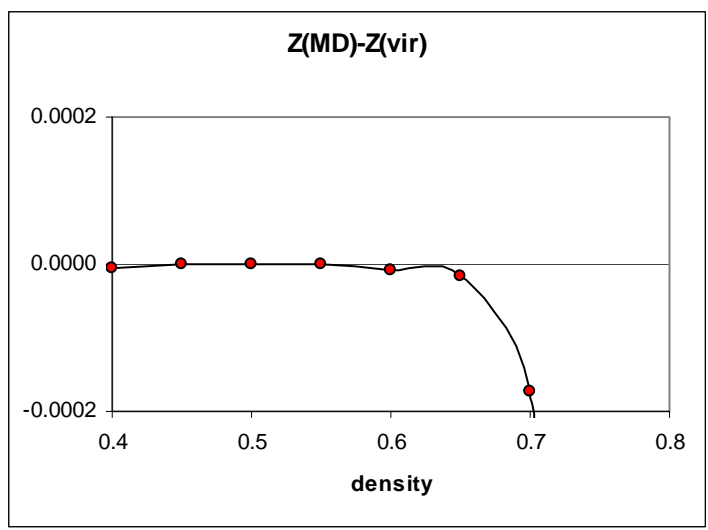

FIG.3 Deviation of the pressure (Z) as given by equation (4) from the computer simulation MD data of reference [1]

The rather abrupt onset of the deviation of equation (4) from the MD data may be indicative of the onset of the first thermodynamic phase transition.  There is presently still some uncertainty regarding the thermodynamic status of the freezing transition in this system. Jaster has recently concluded that there exists a "hexatic mesophase" but no $1^{st}$-order phase transition. Also plotted in FIG.2 are the original coexistence parameters of hard disks determined, with remarkable accuracy considering it was 40 years ago, by Hoover and Ree. The plots show there exists a point on each branch of the respective equations-of-state where the chemical potentials become equal. The state point of Jaster's computation with N = $(1024)^2$ at $\rho$ = 0.918 sits right on the melting point on the crystal branch and hence would not be expected to show 2-phase characteristics.

---

I wish to thank Andrew Masters for bringing to my attention a mistake in reference [1] ($B_{11}$ in 3D is fitted to MD and not calculated by Nathan and Clisby as stated) , and also for bringing  references [2] and [3] to my attention.

# HARD-DISK FLUID

**References**

1. L.V. Woodcock, Ar. Xiv cond-mat/ 0801- 1559 (2008)
2. J. Kolafa and M.. Rottner, Molecular Physics, 104, 3435-3441 (2006)
3. A. Jaster, Physics Letters A330 120-125 (2004)
4. W. G. Hoover and F. H. Ree, J. Chem. Phys. 49 3609-3617 (1968)

**APPENDIX**: Derivation of equation-of-state:

from equations (1) and (3) we have

$$Z = 1 + \sum_{n=2}^{m} B_n \rho^{*(n-1)} + \sum_{n=m+1}^{\infty} (C - An)\rho^{*(n-1)}$$

close first term of summation and differentiate second term

$$Z = 1 + \sum_{n=2}^{m} B_n \rho^{*(n-1)} + C \frac{\rho^{*m}}{(1-\rho^*)} - A \frac{d \sum_{n=m+1}^{\infty} \rho^{*n}}{d\rho^*}$$

replace sum with closure

$$Z = 1 + \sum_{n=2}^{m} B_n \rho^{*(n-1)} + C \frac{\rho^{*m}}{(1-\rho^*)} - A \frac{d \frac{\rho^{*(m+1)}}{(1-\rho^*)}}{d\rho^*}$$

differentiate quotient and simplify

$$Z = 1 + \sum_{n=2}^{m} B_n \rho^{*(n-1)} + C \frac{\rho^{*m}}{(1-\rho^*)} - A \left[ -\frac{m\rho^{*m}}{(1-\rho^*)} - \frac{\rho^{*m}}{(1-\rho^*)^2} \right]$$

or

$$Z = 1 + \sum_{n=2}^{m} B_n \rho^{*(n-1)} + \rho^{*m} \left[ \frac{(C-mA)}{(1-\rho^*)} - \frac{A}{(1-\rho^*)^2} \right]$$

equation (4).